\documentclass[11pt]{article}

\columnwidth\textwidth
\usepackage{graphicx}
\usepackage{epsfig}

\begin{document}
\title{Radial fall of a test particle onto an evaporating
black hole}
\author{Andreas Aste and Dirk Trautmann\\ 
Department for Physics and Astronomy, University of Basel,\\
4056 Basel, Switzerland\\
E-mail: andreas.aste@unibas.ch}
\date{June 8, 2005}

\maketitle
%\twocolumn

\begin{center}
\begin{abstract}
A test particle falling into a classical
black hole crosses the event horizon
and ends up in the singularity within finite eigentime.
In the `more realistic' case of a `classical' evaporating black hole,
an observer falling onto a black hole observes a sudden
evaporation of the hole. This illustrates the fact that the discussion of
the classical process commonly found in the literature may become obsolete
when the black hole has a finite lifetime.
The situation is basically the same for more complex cases,
e.g. where a particle collides with two merging black holes.
It should be pointed out that the model used in this paper is mainly of
academic interest, since the description of the physics near a
black hole horizon still presents a difficult problem which is
not yet fully understood, but our model provides a valuable
possibility for students to enter the interesting field of black hole
physics and to perform numerical calculations of their own which
are not very involved from the computational point of view.\\
\vskip 0.2 cm
\noindent {\bf PACS} numbers: 04.25.-g; 04.70.-s; 04.70.Dy

\end{abstract}
\end{center}

%\newpage
\noindent The Schwarzschild metric generated by an uncharged non-rotating
classical black hole is given by
\begin{equation}
ds^2=\Bigl( 1-\frac{r_s}{r} \Bigr) c^2 dt^2 -\Bigl( 1-\frac{r_s}{r} \Bigr)^{-1}
dr^2-r^2 d \Omega^2, \label{metric}
\end{equation}
where $r_s= 2 G M_0/c^2$ is the Schwarzschild radius, $G$ the gravitational
constant, $M_0$ the mass of the black hole, $c$ the speed of light,
and $d \Omega^2$ denotes the line element squared on a two-dimensional
unit sphere.
Due to the Hawking radiation, the black hole evaporates, if
we exclude any accretion of energy.
For macroscopic black holes, the luminosity is proportional to the
inverse mass squared. The corresponding ansatz for the time-dependent
black hole mass $M(t)$
\begin{equation}
\dot{M}(t) = -k M(t)^{-2},
\end{equation}
leads directly to
\begin{equation}
M(t)^3=\frac{k}{3}(t_0-t), \label{evap}
\end{equation}
where $t_0=3 M_0^3/k$ is the lifetime of the black hole.
Accordingly, we have
\begin{equation}
r_s(t)=k' (t_0-t)^{1/3}. \label{trs}
\end{equation}

It is clear that limited knowledge is available about the
actual evaporation process of black holes at present, which is based
mainly on the classic works of Unruh and Hawking
\cite{Unruh1,Hawking},
and there is an ongoing interest in the subject
\cite{Jacobson,Hooft,Babichev}.
Therefore, our model is purely academic, but we consider it
worthwhile to investigate it due to its simplicity and
because it exhibits the interesting feature that
a particle falling onto a black hole does not vanish
from our universe, since it experiences an immediate
black hole evaporation. A realistic treatment of the
process presented here would involve detailed knowledge
about the physics in the very close vicinity of the
black hole horizon. E.g., Babichev {\em{et al.}} \cite{Babichev}
discussed the case of black holes accreting dark energy in the
form of a phantom energy, where the energy density
$\rho>0$ and the negative pressure $p<0$ fulfill the condition
$\rho c^2+p < 0$. For a black hole surrounded by a constant
phantom energy bath, the black hole mass decreases like
$(t+\tau)^{-1}$, where $\tau>0$ is a characteristic evolution
time. Therefore, we could have chosen decay laws
for our discussion given below
that differ from eq. (\ref{trs}), but the qualitative
results would nevertheless remain the same.
It is also clear that equation (\ref{evap}) will fail to describe
the evaporation process of the black hole in its
final stage.

We also point out that an introduction to the description
of evaporating black holes can be found in a known paper by
William Hiscock \cite{Hiscock}. There, different decreasing functions
$M(t)$ in a slightly different framework using the so-called
Vaidya metric \cite{Vaidya1,Vaidya2} were considered. The main
drawback of using simple approaches like those in \cite{Hiscock}
is the fact that the impact of the
emitted radiation is not taken into account, because the black hole
does not evaporate due to the emission of radiation but by 
artificially adding negative energy to the black hole.

We give here a handwaving argument in order to illustrate
the strange physical conditions that might be expected in the
vicinity of the black hole horizon. For a distant observer,
a massive black hole is a very cold object with Hawking temperature
\begin{equation}
T_{BH}=\frac{\hbar c^3}{4 k_B G M},
\end{equation}
where $k_B$ is the Boltzmann constant.
It is well-known that the condition for thermal equilibrium
in a static system is $\sqrt{g_{00}(\vec{r})} T(\vec{r})=$const.,
where $T(\vec{r})$ is the temperature measured by a local
static observer. This condition can be most easily derived when two systems in
thermal equilibrium are considered, which are coupled only via their
thermal radiation, such that the black body radiation of the two
systems has to travel through the gravitational field.
For the region near the horizon with $r=r_s+\delta$, where
$\delta>0$ is small, $g_{00}$ is given approximately by
\begin{equation}
g_{00}=\Biggl( 1 - \frac{r_s}{r} \Biggr)=
\Biggl( 1 - \frac{r_s}{r_s+\delta} \Biggr) \simeq \frac{\delta}{r_s}.
\end{equation}
Furthermore, the spatial distance $d$ to the horizon is
\begin{equation}
d=\int \limits_{r_s}^{r_s+\delta} \Biggl( 1-\frac{r_s}{r} \Biggr)^{-1/2}
dr
\simeq \int \limits_{0}^{\delta} \sqrt{\frac{r_s}{\delta}} d \delta
=2 \sqrt{r_s \delta}.
\end{equation}
Therefore, a hypothetical static observer very close to the horizon with
$d=l_P=\sqrt{\hbar G/c^3}$
measures a temperature of the order of the Planck temperature $T_P$,
since
\begin{equation}
\frac{k_B T_{BH}}{\sqrt{g_{00}(\delta=l_P^2/4 r_s)}}=
\frac{\hbar c^3}{4 GM} \frac{2 r_s}{l_P}=\frac{\hbar c}{l_P}=k_B T_P.
\end{equation}
For an external
observer, the test particle gets arbitrarily close to the horizon
within finite time, but in contrast to our considerations made
above, it is not in a static state, but accelerated, and does therefore
not experience the extreme Unruh effect as a static observer.
Nevertheless, it will be difficult to make any reasonable predictions about
the fate of a particle approaching the horizon, where the physics
becomes Planckian.

In the following, we assume that the metric
generated by the evaporating black hole is given by expression
(\ref{metric}), but with $r_s$ replaced by the time-dependent
expression given by eq. (\ref{trs}).
In view of the very slow evaporation process and the
low energy density of the emitted Hawking radiation, this is
certainly a satisfactory approximation for our purpose.
Near the black hole horizon, the validity of the approximation
is dubious at best.

From now on, we will use geometric units (G=c=1) throughout.
If we parametrize the (geodesic) worldline of a massive test particle
by its eigentime $\tau$, then
the equation of motion of the particle falling
radially onto the black hole can be derived from the Lagrangian
\begin{equation}
{\cal{L}}=B(r,t) \dot{t}^2 - B(r,t)^{-1} \dot{r}^2, \quad
B(r,t)=\Biggl(1-\frac{k'(t_0-t)^{1/3}}{r} \Biggr),
\end{equation}
where the dot denotes differentiation with respect to the
eigentime $\tau$.
The Euler-Lagrange equations 
\begin{equation}
\partial_\tau \frac{\delta {\cal{L}}}{\delta(\dot{t},\dot{r})}=
\frac{\delta {\cal{L}}}{\delta (r,t)}
\end{equation}
then read
\begin{equation}
\partial_\tau (2 B(r,t) \dot{t})=\partial_t B \, \dot{t}^2
-\partial_t (B^{-1}) \, \dot{r}^2, \label{tdiff}
\end{equation}
\begin{equation}
\partial_\tau (2 B(r,t)^{-1} \, \dot{r})=-\partial_r B \, \dot{t}^2
+\partial_r (B^{-1}) \, \dot{r}^2. \label{rdiff}
\end{equation}
Eq. (\ref{tdiff}) leads to
\begin{equation}
\ddot{r}= \frac{\partial_r B}{2B}  \dot{r}^2 +
\frac{\partial_t B}{B} \dot{r} \dot{t}-
\frac{B}{2} \partial_r B \dot{t}^2, \label{rdotdot}
\end{equation}
\begin{equation}
\ddot{t}=-\frac{\partial_r B}{B} \dot{r} \dot{t}
-\frac{\partial_t B}{2 B} \dot{t}^2-
\frac{\partial_t (B^{-1})}{2B} \, \dot{r}^2.
\end{equation}
Additionally, we can take into account that the four-velocity
of the test particle is constant
\begin{equation}
B \dot{t}^2-B^{-1} \dot{r}^2=1, \label{fourvelocity}
\end{equation}
since we used the eigentime $\tau$ as curve parameter.

We solved eq. (\ref{rdiff}) in conjunction with eq. (\ref{fourvelocity})
numerically. We considered first the analytically solvable
case of a particle falling radially on a stable black hole
(with mass $M$) starting at
rest at a distance $R$
\begin{equation}
\frac{dr}{d \tau} \Biggr|_{r=R}=0, \quad r(\tau=0)=R.
\end{equation}
Introducing the parameter $\eta$ defined by
\begin{equation}
\eta=\arccos \Bigl( \frac{2r}{R}-1 \Bigr),
\end{equation}
one obtains
\begin{displaymath}
r=\frac{R}{2}(1 + \cos \eta), \quad \tau=\sqrt{\frac{R^3}{8M}}
(\eta+\sin \eta),
\end{displaymath}
\begin{equation}
t=2M \sqrt{\frac{R}{2M}-1}
\Bigl[ \eta + \frac{R}{4 M} (\eta+\sin \eta) \Bigr]
+2M \ln \Biggl| \frac{\sqrt{\frac{R}{2M}-1}+\tan \frac{\eta}{2}}
{\sqrt{\frac{R}{2M}-1}-\tan \frac{\eta}{2}} \Biggr| .
\end{equation}
Such a particle reaches the horizon after an eigentime
$\tau_{\mbox{\tiny{H}}}$
\begin{equation}
\tau_{\mbox{\tiny{H}}}= \sqrt{\frac{R^3}{4 r_s}} \Biggl[
\arccos \biggl(\frac{2 r_s}{R}-1 \biggr)+
2\sqrt{\frac{r_s}{R}-\frac{r_s^2}{R^2}} \Biggr]. \label{thorizon}
\end{equation}
This value served as a practical independent check for the accuracy
of our numerical calculation.

A simple explicit finite difference
Euler forward scheme is fully sufficient
for the numerical integration of eq. (\ref{rdotdot}), where
one can eliminate $\dot t$ by using eq. (\ref{fourvelocity}).
{\emph{But}} it is necessary to adapt the eigentime step $\Delta \tau$
during the integration process in order to obtain an
accurate solution. As long as the particle has a relatively
large distance to the horizon, the time step can be held
constant, but in the vicinity of the horizon,
a update of the time step according to
\begin{equation}
\Delta \tau_{n+1}= \frac{\dot{r}_{n-1}}{\dot{r}_{n}}
\Delta \tau_{n}
\end{equation}
for each integration step is a good choice.
The effect is a reduction of the time step when the radial
coordinate $r$ starts to change faster with respect to the
eigentime $\tau$, and the resolution of the trajectory with
respect to the external observer time $t$ is maintained.

Fig. 1 shows the trajectory of the particle falling into
a stable black hole with $r_s=1$ and $R=2$ with respect to
the particle's eigentime $\tau$
and the external observer time $t$.
In this case, the particle reaches the horizon for
$\tau_{\mbox{\tiny{H}}}=\sqrt{2}(\pi/2+1)
\simeq 3.6357$ and the singularity for
$\tau_{\mbox{\tiny{S}}}=\frac{\pi}{2} \sqrt{R^3/r_s} \simeq 4.4429$.

\begin{figure}
        \centering
        \includegraphics[width=9cm]{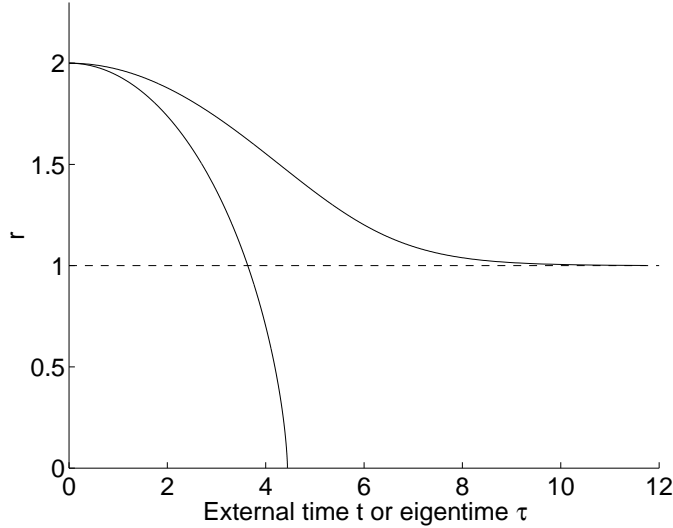}
        \caption{Test particle falling into a stable black hole
        with Schwarzschild radius $r_s=1$. This graph can be found in many standard
        textbooks.}
        \label{fig1}
\end{figure}

\begin{figure}
        \centering
        \includegraphics[width=10cm]{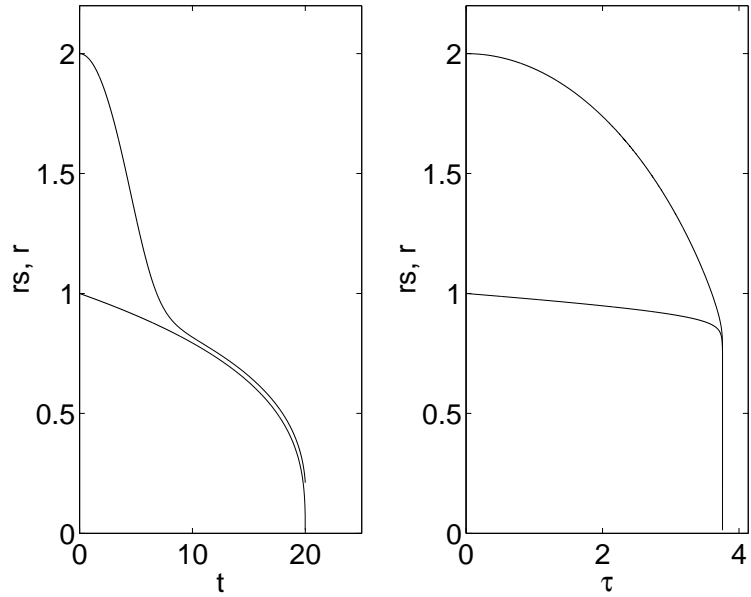}
        \caption{Typical trajectories for a
        test particle falling onto a black hole
        with initial Schwarzschild radius $r_s(0)=1$ and a lifetime
        $t_0$=20.}
        \label{fig2}
\end{figure}

\begin{figure}
        \centering
        \includegraphics[width=9cm]{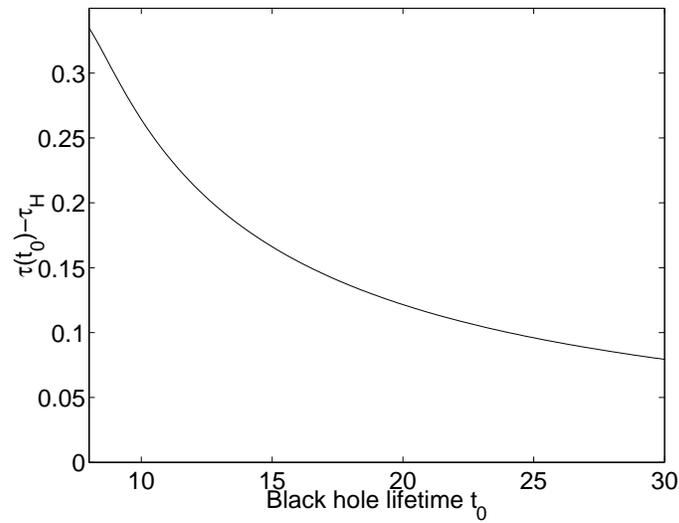}
        \caption{Eigentime $\tau(t_0)$ reduced
         by the eigentime $\tau_{\mbox{\tiny{H}}}$
         needed by a test particle to reach the horizon
         of a stable black hole, versus
         the black hole lifetime $t_0$.}
        \label{fig3}
\end{figure}

Fig. 2 shows the typical situation for an unstable black hole.
In order to obtain reasonable plots, we used 'short' black hole
lifetimes, but the qualitative picture remains the same independently
from the actual lifetime of the black hole which may be
very large for massive holes.
From the external observer's point of view,
the particle impinges on the black hole horizon, where it
remains captured until the black hole evaporates.
In the final evaporation phase,
the particle even looses contact to the horizon due to the fast
decrease of the black hole radius.

From the point of view of the test particle, the black hole
seems to evaporate as soon as the particle approaches the
horizon. The amusing point in our toy model is the fact
that from the 'test particle's point of view', the
evaporation time of the black hole is {\emph{larger}}
for black holes with a short lifetime. This is due to the
fact that the evaporation process
continuously reduces the distance between
the horizon to the particle, such that the particle experiences
a smaller time contraction near the horizon, when the horizon
shrinks faster.
For massive black holes, the black hole evaporation time
approaches the time that the particle would need to reach
the horizon of a stable black hole with $r_s=r_s(\tau=0)$,
given by eq. (\ref{thorizon}).
Fig. 3 shows $\tau(t_0)-\tau_{\mbox{\tiny{H}}}$, i.e. basically
the eigentime of the particle with the same initial conditions
as in Fig. 2 at which the evaporation process of the
black hole with initial lifetime $t_0$ comes to the end. 

We conclude with the remark that within our simplified framework
one should expect basically the same result
for more complex processes, e.g. for a test particle
approaching two merging black holes which evaporate subsequently.
The test particle cannot cross the classical horizon, and an asymptotic observer
will never observe that the particle impinges the horizon within finite time
\cite{Straumann}.

\end{document}